\documentclass[usenatbib]{mn2e}
\usepackage{psfig,graphicx}

\def\chisq{\hbox{$\chi^2$}}
\def\msun{\hbox{${\rm M}_{\odot}$}}
\def\mspy{\hbox{${\rm M}_{\odot}$\,yr$^{-1}$}}
\def\rsun{\hbox{${\rm R}_{\odot}$}}
\def\mstar{\hbox{$M_{\star}$}}
\def\rstar{\hbox{$R_{\star}$}}

\def\sn{\hbox{S/N}}

\def\kms{\hbox{km\,s$^{-1}$}}
\def\vsini{\hbox{$v\sin(i)$}}
\def\vesini{\hbox{$v_{\rm e}\sin(i)$}}

\def\ptt{\hbox{$10^{-4} I_{\rm c}$}}

\def\em{\it}

\def\degr{\hbox{$^\circ$}}
\def\mdot{\hbox{$\dot{m}$}}
\def\ta{\hbox{$t_{\rm A}$}}
\def\ra{\hbox{$R_{\rm A}$}}
\def\vinf{\hbox{$v_{\infty}$}}
\def\Mdot{\hbox{$\dot{M}$}}

\begin{document}

\title[Discovery of a strong magnetic field on HD~191612] {Discovery
of a strong magnetic field on the O star HD~191612: new clues to the
future of $\theta^1$ Orionis C?\thanks{Based on observations obtained at the 
Canada-France-Hawaii Telescope (CFHT) which is operated by the National Research 
Council of Canada, the Institut National des Science de l'Univers of the Centre 
National de la Recherche Scientifique of France, and the University of Hawaii.} }

\makeatletter

\def\newauthor{%
  \end{author@tabular}\par
  \begin{author@tabular}[t]{@{}l@{}}}
\makeatother
 
\author[J.-F.~Donati et al.]
{\vspace{1.5mm}
J.-F.~Donati$^1$, I.D. Howarth$^2$, J.-C. Bouret$^3$, P.~Petit$^1$, C. Catala$^4$, J. Landstreet$^5$ \\ 
$^1$LATT, Obs.\ Midi-Pyr\'en\'ees, 14 Av.\ E.~Belin, F--31400 Toulouse, France
({\tt donati@ast.obs-mip.fr}, {\tt petit@ast.obs-mip.fr}) \\
$^2$ Dept.\ of Physics and Astronomy, University College London,  Gower Street, London WC1E6BT, UK  
({\tt idh@star.ucl.ac.uk}) \\ 
$^3$LAM, Obs.\ de Marseille-Provence, Traverse du Siphon BP 8, F--13376 Marseille Cedex 12, France
({\tt jean-claude.bouret@oamp.fr}) \\ 
$^4$ LESIA, CNRS--UMR 8109, Obs.\ de Paris, 5 Place Janssen, F--92195 Meudon Cedex, France 
({\tt claude.catala@obspm.fr}) \\ 
$^5$ Dept.\ of Physics and Astronomy, University of Western Ontario,  London Ontario N6A3K7, Canada    
({\tt jlandstr@uwo.ca}) 
}

\date{Accepted 2005 October 11.  Received 2005 October 11; in original form 2005 September 23}
\maketitle
 
\begin{abstract} 
From observations made with the ESPaDOnS spectropolarimeter, recently
installed on the 3.6-m Canada--France--Hawaii Telescope, we report the
discovery of a strong magnetic field in the Of?p spectrum variable
HD~191612 -- only the second known magnetic O star (following
$\theta^1$~Ori~C).  The stability of the observed Zeeman signature
over four nights of observation, together with the non-rotational
shape of line profiles, argue that the rotation period of HD~191612 is
significantly longer than the 9-d value previously proposed.  We
suggest that the recently identified 538-d spectral-variability period
is the rotation period, in which case the observed line-of-sight
magnetic field of $-220\pm38$~G implies a large-scale field (assumed
dipolar) with a polar strength of about $-1.5$~kG.  If confirmed, this
scenario suggests that HD~191612 is, essentially, an evolved version
of the near-ZAMS magnetic O star $\theta^1$~Ori~C, but with an even
stronger field (about 15~kG at an age similar to that of
$\theta^1$~Ori~C).  We suggest that the rotation rate of HD~191612,
which is exceptionally slow by accepted O-star standards, could be due to
angular-momentum dissipation through a magnetically confined wind.

\end{abstract}

\begin{keywords} 
stars: magnetic fields --  
stars: winds -- 
stars: rotation -- 
stars: early type -- 
stars: individual:  HD~191612 --
techniques: spectropolarimetry 
\end{keywords}

\section{Introduction} 

Magnetic fields of O stars can strongly impact on the 
physics of the stellar interiors \citep[e.g.,][]{Spruit02} and atmospheres 
\citep[e.g.,][]{Babel97}, and hence on the stars' long-term evolution
\citep[e.g.,][]{Maeder03, Maeder04}.  However, quantifying such effects 
critically depends on our knowledge of the properties of such fields.  

From a theoretical point of view, the origin of magnetic fields in O stars 
is still rather uncertain.  One possibility is that they are fossil 
remnants of the star-formation stage, as proposed for magnetic 
chemically-peculiar Ap and Bp stars \citep[e.g.][]{Mestel99};
if this were the case, we would expect to find about 10\% of O stars with dipolar-type 
magnetic topologies with super-equipartition strengths of at least several 
hundred G.  Another option, gaining recent support among theoreticians, is 
that such fields are produced by dynamo processes, within the convective 
cores \citep{Charbonneau01}, or in a putative subsurface shear layer 
\citep{Macdonald04, Mullan05}; magnetic topologies would then  
depend on rotation rate and feature a significant toroidal component.   

From an observational point of view, however, very few direct constraints exist 
on the strength and topology of magnetic fields in O stars, even though such 
fields are often invoked as a potential explanation for many otherwise enigmatic 
phenomena (e.g., the unexpected properties of the X-ray 
spectra, \citet{Cohen03}).  While several attempts have been made to characterize 
the fields that could be hosted at the surfaces of bright, archetypal O stars such
as $\zeta$~Pup and $\zeta$~Ori, no detections have yet been obtained 
with only one exception: the very young object $\theta^1$~Ori~C \citep{Donati02}.  
One reason for this is that absorption lines of O stars are both relatively 
few in number in the optical, and generally rather broad (due to rotation or 
to some other type of as yet unknown macroscopic mechanism; e.g.,
\citealt{Howarth97}), decreasing dramatically the size of the Zeeman
signatures that their putative fields can induce.  

\begin{table}
\caption[]{Journal of observations.  Columns 1--5 list the
date, heliocentric Julian date, UT time, exposure time, and peak
signal to noise ratio (per 2.6~\kms\ velocity bin) for each
observation.  The last column lists the rms noise level (relative to
the unpolarized continuum level and per 41.6~\kms\ velocity bin) in
the circular polarization profile produced by Least-Squares
Deconvolution (Section~\ref{sec:obs}).}
\begin{tabular}{cccccc}
\hline
Date & HJD          & UT      & $t_{\rm exp}$ & \sn\  & $\sigma_{\rm LSD}$ \\
(2005)     & (2,453,000+) & (h:m:s) &   (s)         &       &   (\ptt) \\
\hline
Jun.\ 22 & 545.02530 & 12:30:55 & $4\times600$ & 340 & 2.6 \\ 
Jun.\ 22 & 545.05519 & 13:13:57 & $4\times600$ & 200 & 4.5 \\ 
Jun.\ 23 & 545.98583 & 11:33:59 & $4\times600$ & 440 & 1.9 \\ 
Jun.\ 23 & 546.01569 & 12:16:59 & $4\times600$ & 450 & 1.8 \\ 
Jun.\ 23 & 546.04559 & 13:00:02 & $4\times600$ & 440 & 1.8 \\ 
Jun.\ 23 & 546.07596 & 13:43:46 & $4\times600$ & 420 & 1.9 \\ 
Jun.\ 24 & 546.97492 & 11:18:11 & $4\times600$ & 440 & 1.9 \\ 
Jun.\ 24 & 547.00479 & 12:01:12 & $4\times600$ & 430 & 1.9 \\ 
Jun.\ 24 & 547.03466 & 12:44:12 & $4\times600$ & 420 & 1.9 \\ 
Jun.\ 25 & 547.98896 & 11:38:19 & $4\times600$ & 280 & 3.0 \\ 
Jun.\ 25 & 548.01933 & 12:22:03 & $4\times600$ & 390 & 2.1 \\ 
Jun.\ 25 & 548.04987 & 13:06:02 & $4\times600$ & 380 & 2.3 \\ 
Jun.\ 25 & 548.08030 & 13:49:50 & $4\times600$ & 330 & 2.9 \\ 
\hline
\end{tabular}
\label{tab:log}
\end{table}

The recent discovery by \citet{Walborn04} that the Of?p star 
HD~191612 varies between spectral types O6 and O8 with an
apparently strict periodicity is significant in this context.  The
H$\alpha$ and He~{\sc i} lines show a particularly strong modulation,
reminiscent of that seen in $\theta^1$~Ori~C \citep{Stahl96};
HD~191612 thus appears at first glance as a very good candidate for
the detection and investigation of a hot-star magnetic field.
However, the derived period of the spectral variability, 538~d, is so
long (by O-star standards) that rotational modulation was considered
unlikely by \citeauthor{Walborn04}.  Instead, they preferred to associate the
spectroscopic variability with an eccentric binary orbit,
hypothesizing that tidally induced non-radial pulsations (NRP) at
periastron may induce periodically enhanced mass loss.
                                                                                                    
However, the lack of any clear orbital radial-velocity variations,
down to a level of a few \kms\ \citep{Walborn03}, is a significant
hurdle to this model, especially since rather large eccentricity is
implied by the NRP hypothesis.  Thus, even though very slow rotation
is at odds with conventional wisdom, the possibility that the observed
variability may be related to rotational modulation of a magnetic star
is by no means excluded.  Motivated by its similarities to
$\theta^1$~Ori~C, we therefore included HD~191612 in a list of
candidates to search for magnetic fields in O-type stars using
ESPaDOnS (Donati et al., in preparation), the new, high-efficiency
spectropolarimeter recently installed on the 3.6-m
Canada--France--Hawaii Telescope (CFHT).

In this paper, we first present the observations and the Zeeman
detection we obtained (Sec.~\ref{sec:obs}).  We then review the
constraints on rotation (Sec.~\ref{sec:rot}), perform simple modelling
of our observations (Sec.~\ref{sec:mod}), and discuss the implications
of our results for rotational evolution of hot stars with strong
radiation-driven winds (Sec.~\ref{sec:disc}).

\section{Observations}
\label{sec:obs}

Spectropolarimetric observations of HD~191612 were collected in 2005
June, as part of a four-night run aimed at investigating the magnetic
fields of hot stars.  (Results obtained on the other stars observed in
the same run will be published separately.)  At the time, the
instrument was suffering a 1.3-mag light loss compared to the optimal
performance obtained during the engineering runs (Donati et al., in
prep.).  This problem was not evident until the time of the run; now
fixed, it turned out to be due to severe damage to the external jacket
of optical fibres linking the polarimeter with the spectrograph
(probably caused during movement of the instrument up and down from
the Cassegrain focus).  Current ESPaDOnS performance is therefore
significantly better than the results given here imply.

\begin{table}
\caption[]{Lines used for Least-Squares Deconvolution.  The 
line depths (column 3) were directly measured from our spectra while 
the Land\'e factors (column 4) were derived assuming LS coupling.  } 
\begin{tabular}{clcc}
\hline
Wavelength & Element & Depth & Land\'e  \\ 
(nm)       &         & ($I_{\rm c}$) & factor       \\
\hline
402.6187   & $\qquad$He {\sc i}  & 0.210 & 1.167 \\
419.9839   & $\qquad$He {\sc ii} & 0.125 & 1.000 \\ 
447.1473   & $\qquad$He {\sc i}  & 0.250 & 1.100 \\  
451.0963   & $\qquad$N {\sc iii} & 0.051 & 1.100 \\  
451.4854   & $\qquad$N {\sc iii} & 0.072 & 1.214 \\  
454.1591   & $\qquad$He {\sc ii} & 0.175 & 1.000 \\   
471.3139   & $\qquad$He {\sc i}  & 0.095 & 1.250 \\  
501.5678   & $\qquad$He {\sc i}  & 0.085 & 1.000 \\  
541.152$\phantom{1}$    & $\qquad$He {\sc ii} & 0.207 & 1.000 \\  
559.2252   & $\qquad$O {\sc iii} & 0.078 & 1.000 \\ 
580.1313   & $\qquad$C {\sc iv}  & 0.172 & 1.167 \\ 
581.1970   & $\qquad$C {\sc iv}  & 0.128 & 1.333 \\  
\hline
\end{tabular}
\label{tab:lines}
\end{table}

HD~191612 was observed each night; altogether, 13
circular-polarization sequences, each consisting of 4 individual
subexposures taken in different polarimeter configurations, were
obtained (see Donati et al., in prep., for details).  All frames were
processed using Libre~ESpRIT (\citealt{Donati97}; Donati et al., in
prep.), a fully automatic reduction package installed at CFHT for
optimal extraction of ESPaDOnS spectra.  The peak signal-to-noise
ratios per 2.6~\kms\ velocity bin range from 200 to 450, depending
mostly on weather conditions (see Table~\ref{tab:log}).

Least-Squares Deconvolution (LSD; \citealt{Donati97}) was applied to
all observations.  This requires the input of a line list, which was
constructed manually to include the few moderate to strong
absorption lines that are present in the spectrum of HD~191612.  As,
essentially, we aim to probe the photosphere of HD~191612, lines
appearing in emission (such as the N~{\sc iii} lines at 463.41 and
464.06~nm) were not included, even though in some cases they may
result from selective emission processes in or near the photosphere.
The strong Balmer lines, all showing clear emission from the wind
and/or circumstellar environment at the time of our observations, were
also excluded from the list, as were other features showing P-Cygni
profiles (such as He~{\sc ii} 468.57~nm and He~{\sc i} 587.56~nm).
This left only 12 usable spectral lines, whose characteristics are
summarized in Table~\ref{tab:lines}.

From those lines we produced a mean circular polarization profile (LSD
Stokes $V$ profile) as well as a mean unpolarized profile (LSD Stokes
$I$ profile) for each spectrum.  Given the breadth of the spectral
lines of HD~191612 (average full width at half depth is
$\sim$150~\kms), the LSD profiles were produced on a spectral grid
with a velocity bin of 41.6~\kms; as this bin size is comparable to
the typical broadening of intrinsic profiles from hot stars, no
information from small-scale magnetic features potentially present at
the surface of the star is expected to be lost in this 
binning process \citep[e.g.][]{Donati02}.  The resulting noise level
in the LSD Stokes $V$ profiles is about 0.02\% of the unpolarized
continuum per 41.6~\kms\ velocity bin (see Table~\ref{tab:log}).

\begin{figure}
\center{\includegraphics[scale=0.32,angle=-90]{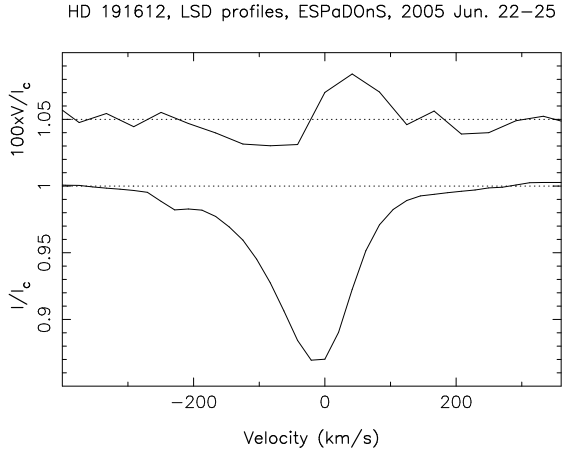}} 
\caption[]{LSD circularly-polarized and unpolarized
profiles of HD~191612 (top, bottom curves respectively) on 2005
June~22--25.  The mean polarization profile is expanded by a factor of
100 and shifted upwards by 1.05 for display purposes.  
}
\label{fig:lsd}
\end{figure}

Averaging together all LSD Stokes $V$ profiles recorded on each of the
four nights of observation (with weights proportional to the inverse
variance of each profile) yields relative noise levels of 2.3, 0.9,
1.1 and 1.2 respectively, in units of \ptt.  Compared to a null-field,
$V=0$ profile, the resulting Stokes $V$ signatures yield
reduced-\chisq\ values of 0.9, 1.9, 1.6 and 1.5, respectively,
implying clear Zeeman detections on each of the last three nights,
though not the first.  However, all four $V$ profiles are mutually 
consistent.  
Averaging all $V$ signatures yields a final noise level of
$0.6 \times 10^{-4} I_{\rm c}$, and a reduced-\chisq\ of 2.7.  We can
thus safely conclude that the star is magnetic; the resulting average
Stokes $V$ profile is shown in Fig.~\ref{fig:lsd} 
and corresponds to a longitudinal field of $-220\pm38$~G.

Note that, even though we excluded lines showing an obvious
contribution from the wind, the LSD Stokes $I$ profile is still
significantly asymmetric (with the blue wing extending to higher
velocities than the red one).  Evidently, most `photospheric' lines
are likely to be affected to a small extent by the stellar wind.  The
Stokes $V$ profile, showing a longer tail at blue wavelengths, is also
slightly impacted by the same effect; however, adding or removing a
few lines in the list does not change the overall shape of the
detected Zeeman signatures (at the noise level), indicating that this
perturbation has no significant consequences for the analysis
presented here.

\section{The rotation period of HD~191612}
\label{sec:rot}

\citet{Walborn04} inferred a rotation period for HD~191612 of
$\sim$9~d, from an estimate of the radius and a measurement of
77~\kms\ for the line-width parameter \vsini.  However, assuming that
the magnetic topology of HD~191612 is essentially dipolar (as found
for all other magnetic OB stars identified to date), a rotation period
of 9~d would imply significant evolution of the observed (projected)
field configuration over a 4-d run, provided that the magnetic dipole
is not aligned with the rotation axis.  We therefore conclude that the
rotation period of the star is most probably significantly longer than
9~d, consistent with the fact that variability is detected in neither
our unpolarized spectra, nor in previous spectroscopic data sets
collected over intervals of a few days \citep[e.g.][]{Walborn03}.

This conclusion is further supported by the obviously non-rotational
shape of the LSD Stokes $I$ profile of HD~191612 (Fig.~\ref{fig:lsd}); 
if `turbulence', of some as yet undetermined physical
nature, dominates the line broadening, then clearly the line-width
parameter \vsini\ significantly overestimates the true projected
equatorial rotation velocity, \vesini\ \citep[cf.][]{Howarth04},
leading to an underestimate of the rotation period.  To examine this
possibility further we compare the observed profile of C$\;${\sc
iv}~580.1~nm to some simple models (Fig.~\ref{fig:rot}).  This line is
rather symmetrical, and is expected to form relatively deep in the
atmosphere, so that it is probably more representative of the
hydrostatic regions than is the LSD $I$ profile.  We took an OStar2002
model profile \citep{Lanz03}, with modest scaling to match the
observed line depth, and applied a rotational convolution with
$\vesini = 77$~\kms.  The resulting profile does indeed match the
observed line {\em width} quite well, but the overal match of the line
shape is very poor.  However, a simple model of gaussian isotropic
turbulence, with {\em zero} rotational broadening, while lacking any
strong physical justification, nonetheless provides an excellent match
to the observations.

Taken together, these arguments strongly suggest a rotation period
significantly longer than the 9-d value previously proposed.  We are
therefore tempted to explore the consequences of identifying the
rotation period of HD~191612 with the 538-d period recently identified
for this star \citep{Walborn04}, on which several spectroscopic
indexes (and in particular H$\alpha$) are observed to vary.  Although
indicators of the implied {\em very} slow rotation are not yet
overwhelming (e.g., constancy of the observed field could
result from a dipole aligned with the rotation axis), the
phenomenological similarities with the young O-type star
$\theta^1$~Ori~C, now strengthened by the magnetic field we have
detected in HD~191612, indicate that the possibility of rotational
modulation merits further consideration.  Thus, while the proposal
clearly needs further scrutiny (e.g., through the monitoring and
potential detection of the rotational modulation of the Zeeman
signature), we nevertheless now explore the consequences of a 538-d
rotation period for HD~191612.

\begin{figure}
\center{\includegraphics[bb=39 42 517 727,scale=0.32,angle=-90]{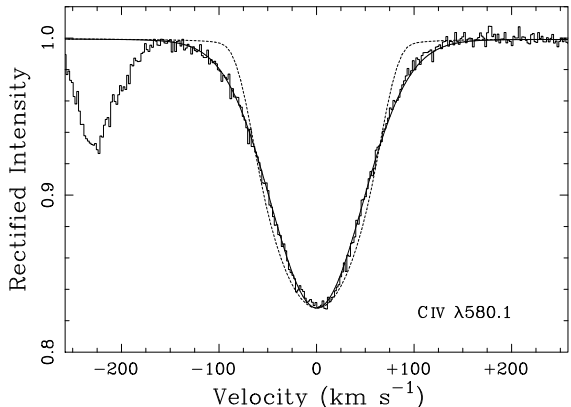}}
\caption[]{The C$\;${\sc iv}~580.1nm profile, compared to models
broadened by rotation ($\vesini=77$~\kms, dashed line) and by
isotropic gaussian turbulence ($\sigma_{\rm V} = 45$~\kms, solid
line).  The feature at $-230$~\kms\ is a diffuse interstellar band.}
\label{fig:rot}
\end{figure}

\section{Modelling the magnetic field of HD~191612}
\label{sec:mod}

We propose that HD~191612, like $\theta^1$~Ori~C, hosts a magnetic
field significantly tilted with respect to the rotation axis, and that
most of the observed variability results from the interaction of the
radiatively driven wind with the stellar magnetic field (as described
in \citealt{Babel97, Donati02, udDoula02, Townsend05, Gagne05}).  In
this framework, the wind coming from each stellar hemisphere is
deflected by the field towards the magnetic equator, where it produces a strong
shock, an X-ray emitting post-shock region (reaching temperatures of
$10^6$ to $10^7$~K), and a cooler and denser disk in the magnetic
equator where the wind plasma piles up before being ejected away from
(or accreted back onto) the star.  
%
%
%
In a generic way, we can then ascribe the observed H$\alpha$ variations,
in particular, to the varying aspect of the cool circumstellar
disk; the emission attains a maximum strength when the disk is
seen pole-on, and a minimum when the disk is seen edge on.  This could
arise through recombination in a moderately optically thick disk.

With this assumption, the longitudinal field of HD~191612 should be maximum
at maximum H$\alpha$ emission, i.e., at phase 0.50 in the ephemeris
obtained by \citet{Walborn04}: ${\rm JD}_{\rm min} = 2,448,315 + 538E$.
Moreover, the strongly reduced H$\alpha$ emission observed around
phase 0.0 indicates that the magnetic equator hosting the H$\alpha$
emitting disk is 
%
%
seen close to edge-on in this viewing configuration.  We therefore
suggest that, as for $\theta^1$~Ori~C, the angle of the magnetic axis
to the rotation axis $\beta$ is close to $90\degr - i$, where $i$ is
the angle of the rotation axis to the line of sight, to ensure that
the magnetic equator is periodically seen edge on (at phase 0.0) by
the observer.  Given the large amplitude of the H$\alpha$ variability, we
can also conclude that neither $i$ nor $\beta$ is likely to be small.

For an initial, schematic, modelling attempt, we therefore propose for
HD~191612 the simplest possible magnetic geometry, with
$i=\beta=45\degr$ (as for $\theta^1$~Ori~C).  In this context, the
magnetic equator is seen edge on at phase 0.0, and the magnetic pole
is facing the observer at phase 0.5, in the \citet{Walborn04}
ephemeris.  Our magnetic observations were taken at phase 0.725; by
fitting the detected Zeeman signature to a magnetic dipole model,
whose single remaining free parameter is the field strength at the
visible pole, we infer that the intensity of the dipole is
$-1.5\pm0.2$~kG.

An alternative is to imagine that the rotation period is actually
twice 538~d, so that the magnetic poles rotate alternately into the
line of sight.  However, we consider this configuration to be
unlikely.  First, it would require $\beta$ to be very close to
90\degr, so that the disk is viewed in the same orientation (and thus
produces the same amount of H$\alpha$ emission) when each pole comes
closest to the observer.  Secondly, this option would produce edge-on
disk viewing episodes that are much shorter than those obtained in the
$\beta=90\degr-i$ case (where the disk approaches only asymptotically
the edge-on configuration), in poor agreement with observations.

\section{Discussion}
\label{sec:disc}


Several observational peculiarities of HD~191612 find a natural explanation 
in the framework of this model.  For instance, \citet{Walborn03} estimate 
that the mass loss of HD~191612 is about 3 times stronger at
phase 0.5 (i.e., when the magnetic pole and associated open field
lines face the observer) than at phase 0.0 (i.e., when the magnetic
equator and associated closed field lines are seen edge on).  This is
precisely what is expected in the context of a magnetically confined
wind (see, e.g., Fig.~6 of \citet{Donati02}).  Similarly, the shape of the 
Hipparcos light curve (\citet{Walborn04}) can be qualitatively explained 
by electron scattering in the disk, redirecting stellar photons to the 
observer at phase 0.5 and away from the line of sight at phase 0.0.  
 
Since a rotation period of 538~d is rather long by O-star standards, 
it naturally raises the question of whether the magnetic
field is responsible for angular-momentum loss that produced the slow
rotation.  Focussing again on the very young star $\theta^1$~Ori~C
suggests clues to answer this question.  Since both stars have similar
masses (of about 40~\msun), HD~191612 can, to first order, be
considered as an evolved version of $\theta^1$~Ori~C (whose age does
not exceed 0.2~Myr).  In this evolution, the radius increases from
about 8~\rsun\ (for $\theta^1$~Ori~C, \citealt{Howarth89, Donati02})
to about 18~\rsun\ (for HD~191612, \citealt{Walborn03}) while the
temperature decreases from about 45,000~K \citep{Howarth89, Donati02}
to about 35,000~K \citep{Walborn03}; this is in rough agreement with
evolutionary models of massive stars \citep[e.g.][]{Schaller92,
Claret04}, from which we then derive an age of about 3--4~Myr for
HD~191612.  This is in good agreement with age estimates for the
Cyg~OB3 association \citep[][e.g.]{Massey95}, of which HD~191612 is a
member \citep{Humphreys78}.

This scenario would imply that HD~191612 hosted a field of about 15~kG
when on the main sequence at an age similar to that of
$\theta^1$~Ori~C today.  However, the corresponding change in the
moment of inertia (about a factor of 3, taking into account the
simultaneous evolution of the fractional gyration radius $k$ from 0.29
to 0.17; \citealt{Claret04}) does not of itself explain the change in
rotation period between $\theta^1$~Ori~C (15~d) and HD~191612 (538~d,
if our model is correct) by more than an order of magnitude.

To investigate whether the magnetic field may be responsible for the
required angular-momentum loss, we can evaluate the magnetic braking
timescale through the simple expression
\begin{equation}
\ta  = k \frac{\mstar}{\mdot} \left( \frac{\rstar}{\ra} \right)^2,
\end{equation}
\noindent where \ra\ is the Alfven radius (i.e., the distance
up to which the wind is magnetically confined) and \mdot\ is the
{\em effective} mass-loss rate (determined by taking into account only the
wind plasma that effectively leaves the star and thus contributes to
angular-momentum loss).  To evaluate the effect of the magnetic field
on the wind of HD~191612, it is useful to consider the wind magnetic
confinement parameter $\eta$, defined by \citet{udDoula02} and
characterizing the ratio between magnetic-field energy density and
the kinetic-energy density of the wind:
\begin{equation}
\eta = B_{\rm eq}^2 \rstar^2 / \Mdot \vinf,
\end{equation}
\noindent where $B_{\rm eq}$ is the equatorial magnetic field,
\Mdot\ is the average mass loss rate, and \vinf\ is the terminal wind
velocity.  If $\eta$ is significantly larger than 1, the magnetic
field confines the wind within the Alfven radius, which
roughly scales as $\eta^{1/4}$ \citep{udDoula02}.

While both HD~191612 and $\theta^1$~Ori~C have similar terminal wind
velocities ($\vinf\simeq2,500$~\kms), HD~191612 exhibits significantly
stronger mass loss (by about an order of magnitude; \citealt{Donati02},
\citealt{Walborn03}).  HD~191612 is also twice as large as, and features a
$\sim1.5\times$ stronger magnetic field than, $\theta^1$~Ori~C,
implying that $\eta$ is similar for both stars ($\eta \sim 10$), and thus
that their Alfven radii are roughly equal (at about 2~\rstar;
\citealt{Donati02}, \citealt{Gagne05}).  It also implies that the effective
mass-loss rate of HD~191612, \mdot, corresponding to plasma evacuated through
field lines opened by the wind (i.e.\ with a magnetic colatitude
smaller than about 45\degr), is about 25\%\ of the actual surface mass
flux, \Mdot, or $\sim1.5\times10^{-6}$~\mspy.  Given the age of
HD~191612, the magnetic braking timescale we derive, of order 1~Myr,
indicates that the field of HD~191612 can, potentially, generate a
strong enough brake to have slowed the star down to the currently
observed rotation rate.

One may argue that the angular-momentum loss of HD~191612 may not have
been as strong as now throughout its past life.  While still on the
main sequence with a radius and mass-loss rate comparable to those of
$\theta^1$~Ori~C, HD~191612 would have had a magnetic field $\sim$15
times larger than that of $\theta^1$~Ori~C, implying that its wind
magnetic confinement parameter was about $\eta \simeq 5000$, and its
Alfven radius reached about 15~\rstar\ at that time.  However, the
correspondingly low effective mass-loss rate (of order
$10^{-8}$~\mspy, or $\sim$5\% of the total surface mass flux) would
still have imposed rotational braking on a timescale of order 1~Myr,
the larger Alfven radius roughly compensating for the smaller
mass-loss rate.  We therefore conclude that the angular-momentum loss
of HD~191612 did not drastically change throughout the life of the
star, and that HD~191612 has had ample time to spin down since it was
born.

This interpretation does not explain why $\theta^1$~Ori~C itself is
apparently rotating more slowly than normal O stars; it is far
too young for its magnetic wind to have  influenced its
rotation rate significantly.  We speculate that, at some stage in its
formation process, the magnetic interaction between the forming star
and its accretion disk may have prevented the star from accumulating as
much angular momentum as normal, weakly magnetic, hot stars, in a
mechanism similar to that proposed for magnetic Ap stars
\citep{Stepien00}.  This speculation needs to be elaborated
properly, however, and tested with adequate observations to see if it can
realistically explain the slow rotation of newly born magnetic O
stars.

Further spectropolarimetric observations of HD~191612, sampling the
538-d period of spectrum variability, are obviously needed to
establish firmly whether this is indeed the rotation period of
HD~191612; to put quantitative constraints on the magnetic-field
geometry of this newly discovered magnetic hot star; and to test the
preliminary conclusions proposed in this paper.  X-ray observations
(such as those already undertaken by Naz\'e et al., in prep.) over the
538-d period will also help to constrain the magnetospheric physics
and geometry, through the fluxes and spectral-line shapes formed in
the postshock hot-plasma torus, as well as from the periodic eclipses
of the torus that the star may naturally provide \citep{Babel97,
Donati02, Gagne05}.

\section*{ACKNOWLEDGMENTS}

We thank the referee, O.~Stahl for valuable comments.


\bibliography{hd191612rv}

\bibliographystyle{mn2e}

\end{document}